\documentclass[12pt]{article}

\usepackage{amsmath}
\usepackage{amssymb}
\usepackage{amsfonts}
\usepackage{lscape}

\begin{document}

\fontsize{12}{6mm}\selectfont
\setlength{\baselineskip}{2em}

$~$\\[.35in]
\newcommand{\dss}{\displaystyle}
\newcommand{\raro}{\rightarrow}
\newcommand{\be}{\begin{equation}}
\newcommand{\ba}{\end{equation}}

\def\sech{\mbox{\rm sech}}
\def\sn{\mbox{\rm sn}}
\def\dn{\mbox{\rm dn}}
\thispagestyle{empty}

\begin{center}
{\Large\bf Quantization of a Particle on a Two-Dimensional } \\ [2mm]   
{\Large\bf Manifold of Constant Curvature  }  \\   [2mm]
\end{center}

\vspace{1cm}
\begin{center}
\noindent
{\bf Paul Bracken}            \\            
{\bf Department of Mathematics,}   \\
{\bf University of Texas,}   \\
{\bf Edinburg, TX  }     \\
{78540}
\end{center}

\vspace{2cm}
\begin{abstract}
The formulation of quantum mechanics on spaces of constant
curvature is studied. It is shown how a transition from a classical
system to the quantum case can be accomplished by the quantization of 
the Noether momenta. These can  be determined by Lie differentiation
of the metric which defines the manifold. For the metric
examined here, it is found that the resulting
Schr\"odinger equation is separable and the
spectrum and eigenfunctions can be investigated in detail.
\end{abstract}

\vspace{1cm}
PACs: 03.65.Ge, 03.65.Ta, 03.65.Aa, 03.65.Ca, 02.40.Hw

\vspace{4mm}

\newpage
\section{Introduction}
\numberwithin{equation}{section}

The study of a quantum particle on a spherical or hyperbolic space
in two-dimensions is amenable to study by treating the scalar curvature
as a parameter. This sort of approach has been of great interest recently {\bf [1,2]}. 
It is certainly physically relevant since
many new phenomenon under investigation occur in two dimensions 
or on two-dimensional manifolds. A very pertinent example of this is the
quantum Hall effect which exhibits the formation of quasiparticles
in the course of its operation {\bf [3]}. The spherical and hyperbolic spaces are
characterized by either a positive or negative  value for the scalar
curvature of the space. In the Euclidean case where the curvature
vanishes, the problem is not complicated because the solutions are
plane-wave states that are in fact momentum eigenfunctions of the
linear momentum operator. Further, here it is the case that plane 
waves are simultaneous eigenfunctions of both the energy and
momentum operators. If the curvature of the space is constant but
different from zero, the canonical momenta do not coincide with
the Noether momenta, so the Noether momenta do not Poisson
commute and as well the quantum versions of these quantities do not
commute as operators. 
These reasons make the situation much more 
complicated in any space with a nonzero or variable curvature.
What is referred to as a plane wave is a Euclidean concept, and it is 
not clear how to generalize the definition to a curved space.
Many of these difficulties can be resolved by adopting a curvature
dependent approach.

Many physical situations can be formulated in terms of an
underlying, curved manifold. In addition to the quantum Hall effect,
the area of quantum dots requires the use of models which are founded
on quantum mechanics on constant curvature spaces.
There is the very active area which studies polynomial billiards, or
systems which are enclosed by geodesic arcs on surfaces with curvature.
Some motions that are integrable in the Euclidean case may become ergodic 
when the curvature of the space becomes negative {\bf [4]}. 
The problem under investigation here
overlaps with the study of quantum chaos in quantum
systems. The entire area of gravitation and cosmology are presently
formulated on a geometric basis. Gravity is a manifestation of the
curvature of the space-time. A space-time is specified or
characterized by defining a metric whose components are used in the
calculation of the curvature of the space-time manifold {\bf [5]}.
There has been a great deal of interest in quantum motion on 
a curved manifold recently. A very different approach to the one 
examined here is to view quantum motion as a submanifold problem
in a generalized Dirac's theory of second-class constraints {\bf [6]}.

It is the objective here to first review some $\kappa$-dependent
formalisms which are appropriate for the description of the 
dynamics on the spaces $M_{\kappa}^2 = ( S_{\kappa}^2, \mathbb E^2,
H_{\kappa}^2 )$ with constant curvature. It is possible to give a
unified approach to both spherical $(\kappa >0)$ and hyperbolic
$( \kappa <0)$ spaces so Euclidean dynamics manifests itself when
the parameter $\kappa =0$. These three spaces can be thought of
as three different cases arising from a family of Riemannian
manifolds $M_{\kappa}^2$ with the curvature appearing as a parameter. 
The components of the metric are selected according to this
geometric structure. The metric of interest here was not
quantized in {\bf [7-9]} or {\bf [10]}. 
Everything can be done in such a way that
applications to other types of system whose Lagrangian can be
defined explicitly in terms of the components of a metric should
be possible.

Once the metric is known, Killing vector fields on the manifold are
determined by means of Lie differentiation of the metric. This procedure
results in a coupled system of partial differential equations in the
unknown component functions of the vector field which can be easily solved.
A Hilbert space can be constructed for the problem by
defining a measure which is annihilated upon Lie differentiation with
respect to this set of linearly independent vector fields.
The Killing vector fields provide the Noether momenta for the system.
The Hamiltonian is obtained by means of the usual canonical 
transformation from the Lagrangian and subsequently written in terms
of these momenta. The quantization algorithm can then be applied
to the components of the Noether momenta which appear in the 
Hamiltonian. Thus, the Hamiltonian can be quantized in this way.
Once the Hamiltonian has been given, the Schr\"odinger
equation can be written down. It is remarkable to note that for the
metric which is defined and used here, the Schr\"odinger equation
is separable, and moreover, the energy and wavefunctions can be 
calculated from it in closed form.

\section{Metric and Associated Hamiltonian}

From the geometric point of view, the sphere $S_{\kappa}$, the
Euclidean plane $\mathbb E^2$ and the hyperbolic plane $H_{\kappa}^2$
represent three different objects which make up a family of Riemannian
manifolds. These are grouped together and referred 
to as $M_{\kappa}^2 = ( S_{\kappa}^2, \mathbb E^2,
H_{\kappa}^2)$ of curvature $\kappa$. Suppose a general metric is
assigned to this class of spaces by making use of the $\kappa$-dependent
trigonometric and hyperbolic functions defined as
\be
S_{\kappa} (x) =
\begin{cases}
\dss\frac{1}{\sqrt{\kappa}} \sin ( \sqrt{\kappa} x),  &\text{$\kappa > 0$,}\\
x,   &\text{$\kappa =0$,}\\
\dss\frac{1}{\sqrt{\kappa}} \sinh ( \sqrt{\kappa} x),   &\text{$\kappa < 0$.}\\
\end{cases}
\label{eqII1}
\ba
From \eqref{eqII1}, the functions $C_{\kappa} (x)$ and $T_{\kappa} (x)$ can be
defined
\be
C_{\kappa} (x) = \frac{d S_{\kappa} (x)}{dx},
\qquad
T_{\kappa} (x) = \frac{S_{\kappa} (x)}{C_{\kappa} (x)}.
\label{eqII2}
\ba
A general metric in geodesic polar coordinates $( \rho, \varphi )$ on
$M_{\kappa}^2$ is defined in the following way,
\be
g = d \rho \otimes d \rho + S_{\kappa}^2 ( \rho) \, d \varphi \otimes d \varphi.
\label{eqII3}
\ba
Several Lagrangians can be obtained from \eqref{eqII3} by diffeomorphisms
and can be considered to be dynamically equivalent at the classical level.
One of these Lagrangians, which has not been examined before,
will be used as the starting point for the
construction of the Hamiltonian quantum system. 
Let $M$ be a Riemannian or pseudo-Riemannian
manifold whose metric evaluated at a point $p \in M$ is $g(p)$. On the 
tangent space $TM$, a Lagrangian can be defined by first giving the
kinetic energy in terms of the components of the metric {\bf [11]}
\be
T = \frac{1}{2}  g_{ij} v^i v^j.
\label{eqII4}
\ba
The Lagrangian of geodesic motion which corresponds to \eqref{eqII3}
on $M_{\kappa}^2$ given by \eqref{eqII4} plus a potential function is
\be
L ( \kappa) = \frac{1}{2} ( v_{\rho}^2 + S_{\kappa}^2 ( \rho) v_{\varphi}^2 )
+ V(r).
\label{eqII5}
\ba
Several diffeomorphic versions of \eqref{eqII5} can be presented.
Consider a $\kappa$-dependent transformation defined by $\rho \raro r' =
T_{\kappa} ( \rho)$. This transformation puts the metric \eqref{eqII3}
into the form
\be
g = \frac{1}{( 1 + \kappa r^2)^2} \, dr \otimes dr 
+ \frac{r^2}{1 + \kappa r^2} \, d \varphi \otimes d \varphi,
\label{eqII6}
\ba
after writing $r$ in place of $r'$,
and transforms takes Lagrangian \eqref{eqII5} into the form,
\be
L_H ( \kappa) = \frac{1}{2} ( \frac{v_r^2}{( 1 + \kappa r^2)^2}
+ \frac{r^2}{1 + \kappa r^2} v_{\varphi}^2 ) + V(r).
\label{eqII7}
\ba
It is worth mentioning that \eqref{eqII7} can be transformed into
Cartesian form by means of the following relation,
\be
v_x^2 + v_y^2 + \kappa ( x v_y - y v_x)^2
= v_r^2 + r^2 ( 1 + \kappa r^2) v_{\varphi}^2.
\label{eqII8}
\ba
Consequently, Lagrangian \eqref{eqII7} is given by
\be
L_H (\kappa ) = \frac{1}{2 ( 1 + \kappa r^2)^2}
 [ v_x^2 + v_y^2 + \kappa ( x v_y - y v_x )^2 ] - \frac{1}{2} \alpha^2 r^2.
\label{eqII9}
\ba
where a potential which depends on $r$ has been included and $r^2 = x^2 + y^2$.

Setting $v_x = \dot{x}$ and $v_y = \dot{y}$, the canonical
momenta are determined to be
\be
p_x = \frac{\partial L}{\partial v_x} = \frac{1}{(1 + \kappa r^2)^2} 
[ v_x - \kappa \, y \,( x v_y - y v_x)  ],
\qquad
p_y = \frac{\partial L}{\partial v_y} = \frac{1}{(1 + \kappa r^2)^2}
[ v_y - \kappa \, x \, (x v_y - y v_x) ].
\label{eqII10}
\ba
These are required in order to compute the classical Hamiltonian.
Solving \eqref{eqII10} for $v_x$ and $v_y$, it is found that
\be
v_x = (1 + \kappa r^2) ((1 + \kappa x^2) p_x + \kappa x y p_y),
\qquad
v_y = (1 + \kappa r^2) (( 1 + \kappa y^2) p_y + \kappa x y p_x ).
\label{eqII11}
\ba
The Hamiltonian in the $(x,y)$ coordinates is obtained by means
of the usual canonical transformation
\be
H (\kappa) = p_x v_x + p_y v_y - L_H (\kappa)
= \frac{1}{2} ( 1 + \kappa r^2) \{ p_x^2 + p_y^2 + \kappa
( y p_y + x p_x)^2 \} + \frac{1}{2} \alpha^2 (x^2 + y^2).
\label{eqII12}
\ba
The Hamiltonian is simpler and more useful in cylindrical coordinates
and it is written in this form now.
Introducing $v_r = \dot{r}$ and $v_{\varphi}= \dot{\varphi}$,
the canonical momenta in the cylindrical variables are 
\be
p_r = \frac{\partial L}{\partial v_r} = \frac{\dot{r}}{(1 + \kappa r^2)^2},
\qquad
p_{\varphi} = \frac{\partial L}{\partial v_{\varphi}} = \frac{r^2}
{1 + \kappa r^2} \dot{\varphi}.
\label{eqII13}
\ba
Solving \eqref{eqII13} for $v_r$ and $v_{\varphi}$, it is found that
\be
v_r = (1 + \kappa r^2)^2 p_r,   
\qquad
v_{\varphi} = \frac{1}{r^2} ( 1 + \kappa r^2 )  p_{\varphi}.
\label{eqII14}
\ba
The Hamiltonian in terms of cylindrical variables is then calculated to be
\be
H (\kappa) = p_r v_r + p_{\varphi} v_{\varphi} - L_H (\kappa)
= \frac{1}{2} ( 1 + \kappa r^2)^2 p_r^2 + \frac{1}{2 r^2} ( 1 + \kappa r^2) p_{\varphi}^2 
+ \frac{1}{2} \alpha^2 r^2.
\label{eqII15}
\ba
All of the calculations given here are easy to verify by means
of symbolic manipulation {\bf [12]}.

\section{Noether Symmetries}

A set of three linearly independent Killing vector fields will be
calculated for the metric presented in \eqref{eqII6}.
If for a certain vector field the Lie derivative of the metric
vanishes, this vector field is called a Killing vector field. This 
type of vector field can be thought of as an infinitesimal
generator of isometries of the $\kappa$-dependent metric \eqref{eqII6}.

Let $X$ be a vector field in terms of the cylindrical $(r, \varphi)$
coordinates defined by
\be
X = f (r, \varphi) \frac{\partial}{\partial r}
+ h ( r, \varphi) \frac{\partial}{\partial \varphi}.
\label{eqIII1}
\ba
The two functions $f$ and $h$ will be determined in such
a manner that the Lie derivative of $g$ vanishes,
\be
{\cal L}_X \, g =0.
\label{eqIII2}
\ba
Differentiating the metric with respect to $X$,
it is found that
$$
{\cal L}_X \, g = f \frac{\partial}{\partial r}
( \frac{1}{(1 + \kappa r^2)^2} ) \, dr \otimes dr
+ \frac{1}{(1 + \kappa r^2)^2}
( \frac{\partial f}{\partial r} dr \otimes dr
+ \frac{\partial f}{\partial \varphi} \, d \varphi \otimes dr
+ \frac{\partial f}{\partial r} \, dr \otimes dr
+ \frac{\partial f}{\partial \varphi} \, dr \otimes d \varphi)
$$
$$
+ f \frac{\partial}{\partial r} ( \frac{r^2}{1 + \kappa r^2} )
d \varphi \otimes d \varphi + \frac{r^2}{1 + \kappa r^2}
( \frac{\partial h}{\partial r} \, dr \otimes d \varphi
+ \frac{\partial h}{\partial \varphi} \, d \varphi \otimes d \varphi
+ \frac{\partial h}{\partial r} \, d \varphi \otimes d r + 
\frac{\partial h}{\partial \varphi} \, d \varphi \otimes d \varphi ).
$$
Collecting like terms, it is required that the coefficient of
each tensor product in this expression must vanish for \eqref{eqIII2}
to hold. This produces the following system of three coupled first order
partial differential equations in terms of $f$ and $h$,
\be
\frac{\partial f}{\partial r} - \frac{2 \kappa r}{1 + \kappa r^2} f =0,
\quad
\frac{\partial f}{\partial \varphi} + r^2 (1 + \kappa r^2)
\frac{\partial h}{\partial r} =0,
\quad
r (1 + \kappa r^2) \frac{\partial h}{\partial \varphi} + f =0.
\label{eqIII3}
\ba
This system of equations can be readily solved to yield the
following general solution for $f$ and $h$,
\be
f (r, \varphi) = (1 + \kappa r^2) ( C_1 \sin \varphi + C_2 \cos \varphi ),
\qquad
h (r, \varphi ) = \frac{1}{r} ( C_1 \cos \varphi - C_2 \sin \varphi )
+ C_3.
\label{eqIII4}
\ba
The three required independent vector fields can be specified by
choosing the constants appropriately; for example,
$(C_1, C_2, C_3) = (1,0,0), (0,1,0), (0,0,1)$. For this choice, we have 
the following vector fields,
\be
X_1 = (1 + \kappa r^2) \cos \varphi \frac{\partial}{\partial r}
- \frac{1}{r} \sin \varphi \frac{\partial}{\partial \varphi},
\qquad
X_2 = (1 + \kappa r^2) \sin \varphi \frac{\partial}{\partial r}
+ \frac{1}{r} \cos \varphi \frac{\partial}{\partial \varphi},
\qquad
X_J = \frac{\partial}{\partial \varphi}.
\label{eqIII5}
\ba
The Lie commutator brackets for \eqref{eqIII5} can be calculated
and are given by
\be
[ X_1, X_2 ] =- \kappa X_J,
\qquad
[X_1, X_J ] = X_2,
\qquad
[ X_2, X_J ] =- X_1.
\label{eqIII6}
\ba
The set of vector fields given by \eqref{eqIII5} are the required
Noether symmetries, so the coefficient functions satisfy system
\eqref{eqIII3} and the associated constants of the motion are
\be
P_1 = ( 1 + \kappa r^2) \cos \varphi \, p_r - \frac{1}{r} \sin \varphi \, p_{\varphi},
\qquad
P_2 =( 1 + \kappa r^2) \sin \varphi \, p_r + \frac{1}{r} \cos \varphi \, p_{\varphi},
\qquad
J = p_{\varphi}.
\label{eqIII7}
\ba
The classical Poisson bracket of two dynamical quantities $F$ and $G$ is
defined by
\be
\{ F, G \} = \frac{\partial F}{\partial r} \frac{\partial G}{\partial p_r}
+ \frac{\partial F}{\partial \varphi} \frac{\partial G}{\partial p_{\varphi}}
- \frac{\partial F}{\partial p_r} \frac{\partial G}{\partial r}
- \frac{\partial F}{\partial p_{\varphi}} \frac{\partial G}{\partial \varphi}.
\label{eqIII8}
\ba
For the case in which the variables $F$, $G$ are replaced by
$P_1$, $P_2$ and $J$ in \eqref{eqIII8}, the following brackets are
obtained
\be
\{ P_1, P_2 \} = \kappa J,
\qquad
\{ P_1, J \} = - P_2,
\qquad
\{ P_2, J \} = P_1.
\label{eqIII9}
\ba
Using Hamiltonian \eqref{eqII15}, the following Poisson 
brackets are also found
\be
\{ P_1, H \} =0,  
\qquad
\{ P_2, H \} =0,
\qquad
\{ J, H \} =0.
\label{eqIII10}
\ba
At the classical level, it is clear using \eqref{eqIII7} that
\be
P_1^2 + P_2^2 + \kappa J^2  = ( 1 + \kappa r^2)^2 p_r^2 
+ \frac{1}{r^2} ( 1 + \kappa r^2) p_{\varphi}^2.
\label{eqIII11}
\ba
The classical Hamiltonian in terms of the variables
$P_1$, $P_2$ and $J$ including a mass $m$ is given by
\be
H = \frac{1}{2 m} [ P_1^2 + P_2^2 + \kappa J^2] + \frac{1}{2} \alpha^2 r^2.
\label{eqIII12}
\ba
The only measure on the space $\mathbb R^2$ that is invariant under the
action of the vector fields \eqref{eqIII6} in the sense that the Lie 
derivative vanishes should be used to construct the Hilbert space. 
By starting with a function of $r$ times $dr \wedge d \varphi$,
the function can be determined by differentiating with respect to
these vector fields
\be
{\cal L}_{X_i} \, d \mu_{\kappa} =0,
\qquad i=1,2, J.
\label{eqIII13}
\ba
An ordinary differential equation results which can be solved to give the measure as
\be
d \mu_{\kappa} = \frac{r}{(1 + \kappa r^2)^{3/2}} \, dr \wedge d \varphi.
\label{eqIII14}
\ba
Thus, the space carries a measure somewhat different from the one
in {\bf [10]}. The quantum Hamiltonian would be self-adjoint, not in the
standard space $L^2 (\mathbb R^2)$, but in the Hilbert space $L^2 (d \mu_{\kappa})$. 
In the spherical case, the space is $L^2 ( \mathbb R^2, d \mu_{\kappa})$,
and in the hyperbolic case, it is $L^2_0 ( \mathbb R^2_{\kappa}, d \mu_{\kappa})$,
where $\mathbb R^2_{\kappa}$ denotes the region $r^2 \leq 1/ \kappa$, and
functions vanish at the boundary of the region.

\section{Quantization of the Hamiltonian and Schr\"odinger Equation}

A procedure which allows the Hamiltonian of the model to be
quantized can be formulated based on property \eqref{eqIII13} of the
measure. The idea is to consider functions and linear operators 
which are defined on a related space. This space can be defined by taking the 
two-dimensional real plane and using the measure \eqref{eqIII14} on it.
The quantum operators which will be defined represent the
quantum version of the Noether momenta. They must be self-adjoint in the
space $L^2 ( \mathbb R, d \mu_{\kappa} )$.

The transition from classical to quantum mechanics by means of the
Noether momenta \eqref{eqIII7} is now represented by the following
correspondence
$$
P_1 \raro \hat{P}_1 =- i \hbar \{ ( 1 + \kappa r^2) \cos \varphi 
\frac{\partial}{\partial r} - \frac{1}{r} \sin \varphi \frac{\partial}
{\partial \varphi} \},
$$
\be
P_2 \raro \hat{P}_2 =-i \hbar \{ (1 + \kappa r^2) \sin \varphi 
\frac{\partial}{\partial r} + \frac{1}{r} \cos \varphi 
\frac{\partial}{\partial \varphi} \},
\label{eqIV1}
\ba
$$
J \raro \hat{J} = - i \hbar \frac{\partial}{\partial \varphi}.
$$
Under transformation \eqref{eqIV1}, the classical Hamiltonian 
\eqref{eqIII12} is transformed into the following operator,
\be
\hat{H} = - \frac{\hbar^2}{2m} (1 + \kappa r^2 )
[ ( 1 + \kappa r^2) \frac{\partial^2}{\partial r^2}
+ (1 + 2 \kappa r^2) \frac{1}{r} \frac{\partial}{\partial r}
+ \frac{1}{r^2} \frac{\partial^2}{\partial \varphi^2} ] +
\frac{1}{2} \alpha^2 r^2.
\label{eqIV2}
\ba
The Hamiltonian \eqref{eqIV2} immediately yields the Schr\"odinger equation,
\be
- \frac{\hbar^2}{2m} (1 + \kappa r^2)
[ ( 1 + \kappa r^2) \frac{\partial^2}{\partial r^2}
+( 1 + 2 \kappa r^2) \frac{1}{r} \frac{\partial}{\partial r}
+ \frac{1}{r^2} \frac{\partial^2}{\partial \varphi^2} ] \Psi
+ \frac{1}{2} \alpha^2 r^2 \Psi = E \Psi.
\label{eqIV3}
\ba
To solve \eqref{eqIV3}, it is advantageous to have \eqref{eqIV3} in a form 
in which the physical constants have been scaled out of the equation.
Introduce the new constant $\alpha = \sqrt{m} \beta$ into
the equation so that
\be
- \frac{\hbar^2}{2m} ( 1 + \kappa r^2) [ (1 + \kappa r^2)
\frac{\partial^2}{\partial r^2} + (1 + 2 \kappa r^2) 
\frac{1}{r} \frac{\partial }{\partial r} + \frac{1}{r^2}
\frac{\partial^2}{\partial \varphi^2} ] \Psi + \frac{1}{2}
m \beta^2 r^2 \Psi = E \Psi.
\label{eqIV4}
\ba
Now introduce the following set of new variables 
$( \bar{r}, \bar{\kappa}, {\cal E})$ which are defined to be
\be
r = \sqrt{\frac{\hbar}{m \beta}} \bar{r},
\qquad
\kappa = \frac{m \beta}{\hbar} \bar{\kappa},
\qquad
E = \hbar \beta {\cal E}.
\label{eqIV5}
\ba
Consequently, $\kappa r^2 = \bar{\kappa} \bar{r}^2$ and upon substituting 
\eqref{eqIV5} into Schr\"odinger equation \eqref{eqIV4}, it transforms into
$$
- \frac{\hbar^2}{2m} ( 1 + \bar{\kappa} \bar{r}^2 ) \frac{m \beta}{\hbar}
\frac{\partial^2}{\partial \bar{r}^2} + ( 1 + 2 \bar{\kappa} \bar{r}^2)
\frac{m \beta}{\hbar} \frac{1}{\bar{r}} \frac{\partial}{\partial \bar{r}}
+ \frac{m \beta}{\hbar} \frac{1}{\bar{r}^2} \frac{\partial^2}{\partial \varphi^2} ]
\Psi + \frac{1}{2} m \beta^2 \frac{\hbar}{m \beta} \bar{r}^2 \Psi = E \Psi.
$$
Removing the physical constants from the equation and dropping the bars
from the variables, we obtain
\be
(1 + \kappa r^2) [ ( 1 + \kappa r^2 ) \frac{\partial^2}{\partial r^2}
+ ( 1 + 2 \kappa r^2) \frac{1}{r} \frac{\partial}{\partial r} +\frac{1}{r^2}
\frac{\partial^2}{\partial \varphi^2} ] \Psi - r^2 \Psi + 2 {\cal E} \Psi =0.
\label{eqIV6}
\ba

\section{Energies and Wavefunctions for the Model.}

It is particularly interesting to observe that the spectrum and the
structure of the wave functions for this Hamiltonian can be
investigated for the system defined by metric \eqref{eqII6}.
This is mainly due to the fact that the Schr\"odinger equation
\eqref{eqIV6} is separable. It will be shown that there exist
solutions to it of the form,
\be
\Psi ( r, \varphi ) = R(r) \Phi (\varphi),
\label{eqV1}
\ba
where $R$ and $\Phi$ are functions of the variables $r$ and
$\varphi$, respectively. Substitute \eqref{eqV1} into
Schr\"odinger equation \eqref{eqIV6} so it takes the form
\be
\Phi (\varphi ) (1 + \kappa r^2) \{ ( 1 + \kappa r^2) R'' + (1 + 2 \kappa r^2)
\frac{R'}{r} \} + \frac{1}{r^2} ( 1 + \kappa r^2) R \ddot{\Phi}
- r^2 R \Phi + 2 {\cal E} R \Phi =0.
\label{eqV2}
\ba
It is possible to separate this equation by first introducing
a separation constant $\beta$. Equation \eqref{eqV2} takes the form,
\be
\frac{r^2}{R} [ ( 1 + \kappa r^2) R'' + (1 + 2 \kappa r^2) \frac{R'}{r} ]
- \frac{r^4}{1 + \kappa r^2} + 2 \frac{ r^2}{1 + \kappa r^2} {\cal E}
=- \frac{\ddot{\Phi}}{\Phi} = \beta^2.
\label{eqV3}
\ba
This result is equivalent to the following pair of equations for
$\Phi$ and $R$,
\be
\begin{array}{c}
\ddot{\Phi} + \beta^2 \Phi =0,   \\
\\
r^2 ( 1 + \kappa r^2) R'' + r ( 1 + 2 \kappa r^2) R' 
- \dss\frac{r^4}{1 + \kappa r^2} R + \dss\frac{2 r^2}{1 + \kappa r^2}
{\cal E} R - \beta^2 R =0.   \\
\end{array}
\label{eqV4}
\ba
The equation for $\Phi$ has the exponential solutions of the form
$$
\Phi ( \varphi) = e^{\pm i \beta \varphi}.
$$ 
The parameter $\kappa$ is relegated to the radial equation. The radial
solution factorizes to the form,
\be
R ( r, \kappa ) = F ( r, \kappa) ( 1 + \kappa r^2)^s.
\label{eqV5}
\ba
The radial equation then becomes an equation satisfied by the function
$F (r) = F ( r, \kappa)$,
\be
r^2 ( 1 + \kappa r^2)^2 \, F'' (r) + r ( 1 + \kappa r^2)( 2 \kappa (1+2s)
r^2 +1) F' (r) +( 2 s (1+s) \kappa^2 -1) r^4 + ( 4 \kappa s - \kappa \beta^2 
+ 2 {\cal E} ) r^2 - \beta^2 )F =0.
\label{eqV6}
\ba
To solve this, a specific value for the parameter $s$ is taken.
Consider the case in which $s$ is given by
\be
s = \frac{1}{4} - \frac{q (\kappa)}{4 \kappa},
\qquad
q= q ( \kappa) = \sqrt{ \kappa^2 +8 \kappa {\cal E} +4}.
\label{eqV7}
\ba
In this instance, it follows that in the small $\kappa$ limit,
$$
\lim_{\kappa \raro 0} \, R( r, \kappa ) = F (r) e^{-r^2 /2}.
$$
This choice for $s$ gives an equation in which the 
$r^4$ dependence has disappeared from the last term
of \eqref{eqV6} and it becomes,
\be
r^2 ( 1 + \kappa r^2) F'' (r) + (( 3 \kappa - q )r^2 +1) r F' (r) + 
(( 2 {\cal E} + \kappa -q) r^2 - \beta^2 ) F (r) =0.
\label{eqV8}
\ba
The indicial equation for \eqref{eqV8} implies that there is a 
regular solution at $r=0$ of the form,
$$
F (r) = r^{\beta} f (r).
$$
Substituting this form into \eqref{eqV8}, it is found that
$f (r)$ must satisfy,
\be
r  (1 + \kappa r^2) f^{''} (r) + ( 1 + 2 \beta + ( 3 \kappa +
2 \beta \kappa -q) r^2 ) \, f' (r) + r (\kappa ( \beta +1)^2 -q ( \beta +1) 
+ 2 {\cal E} ) f (r) =0.
\label{eqV9}
\ba
In the Euclidean case, the curvature scalar $\kappa =0$, and this equation
reduces to
\be
r f'' (r) + ( 2 \beta + 1 - 2 r^2) f' (r) -2 ( 1 + \beta - {\cal E})
r f(r) =0.
\label{eqV10}
\ba
The solution which is regular at $r=0$ is the Kummer-M function
\be
f (r) = c_0 K_M ( a ; c; r^2),
\label{eqV11}
\ba
where $c_0$ is a constant and the parameters $a$ and $c$ are
defined by
\be
a = \frac{1}{2} ( 1 + \beta - {\cal E}),
\qquad
c = \beta^2 +1.
\label{eqV12}
\ba
The physically acceptable solutions are the polynomial solutions that
appear when $a =- n_r$, $n_r =0,1,2, \cdots$. This choice gives rise to 
a quantization condition on the energy spectrum.

To obtain a recursion relation for the coefficients $a_n (\kappa )$, 
write $f (r)$ in the form of a power series,
\be
f (r) = \sum_{n=0}^{\infty} \, a_n (\kappa) r^n.
\label{eqV13}
\ba
This function is substituted into \eqref{eqV9}, and the required recursion 
relation is determined to be, $a (\kappa )=0$ and,
\be
a_{n+1} ( \kappa ) =
\frac{(n + \beta ) q - (n + \beta )^2 \kappa - 2 {\cal E}}
{ (n+1)(n +2 \beta +1 )} 
a_{n-1} ( \kappa), 
\quad
n=1,2, \cdots.
\label{eqV14}
\ba
Since $a_0 (\kappa)$ does not depend on $r$, \eqref{eqV14} implies that
\eqref{eqV13} is made up of even powers of $r$. A form for $f (r)$ can 
also be obtained in terms of the hypergeometric function by putting the
equation in hypergeometric form. Introduce then the variable $t= r^2$
so that \eqref{eqV9} becomes
\be
t ( 1 + \kappa t) f_{tt} + ( \beta +1 +( 2 \kappa + \beta \kappa - \frac{q}{2} ) \, t )
f_t + \frac{1}{4} (( \beta + 1 )^2 \kappa - q ( \beta +1 ) + 2 {\cal E} ) f =0.
\label{eqV15}
\ba
For the last step, introduce the new variable $s=- \kappa t$ so the
equation becomes,
\be
s (1 -s ) f_{ss} + ( \beta + 1 - \frac{1}{2 \kappa} ( 2 \kappa ( 2 + \beta ) -q) s )
f_s - \frac{1}{ 4 \kappa} ( \kappa ( \beta +1)^2 - ( \beta +1) q +
2 {\cal E}) f = 0.
\label{eqV16}
\ba
This equation is exactly the Gauss hypergeometric equation,
\be
s (1 -s ) f_{ss} + (c - ( 1 + a_{\kappa} + b_{\kappa} ) s ) \, f_s -
a_{\kappa} b_{\kappa} f =0,
\label{eqV17}
\ba
where the constants are defined as
\be
c = \beta + 1,
\quad
a_{\kappa} + b_{\kappa} = \frac{2 ( \beta + 1 ) \kappa - q}{ 2 \kappa},
\quad
a_{\kappa} b_{\kappa} = \frac{1}{4 \kappa} (( \beta + 1)^2 \kappa
- ( \beta + 1 ) q + 2 {\cal E}).
\label{eqV18}
\ba
The solution of \eqref{eqV16} for $f (r)$ which is regular at $r=0$ can 
be expressed in terms of the generalized hypergeometric function,
\be
f (r) = \, _2 F_1 ( a_{\kappa}; b_{\kappa}; c ; r^2),
\label{eqV19}
\ba
where $a_{\kappa}$ and $b_{\kappa}$ are given by
\be
a_{\kappa} = \frac{ 2 ( 1 - \beta) \kappa - \sqrt{\kappa^2 +4} -q}{4 \kappa},
\quad
b_{\kappa} = \frac{2 ( 1 - \beta ) \kappa + \sqrt{\kappa^2 +4} -q}{4 \kappa},
\quad
c = \beta +1.
\label{eq20}
\ba
Physically acceptable solutions which are determined as eigenfunctions of 
the singular $\kappa$-dependent Sturm-Liouville problem appear when one of 
the two $\kappa$-dependent coefficients $a_{\kappa}$ or $b_{\kappa}$
coincides with zero or a negative integer,
\be
a_{\kappa} = - N_r,
\qquad
b_{\kappa} =- N_r,
\qquad
N_r =0,1,2, \cdots.
\label{eqV21}
\ba
This restricts the energy to one of the following values
\be
{\cal E} ( \kappa) = \frac{1}{2} ( 2 N_r + \beta + 1 )(( 2 N_r + \beta +1)
\kappa - \sqrt{ \kappa^2 + 4} ).
\label{eqV22}
\ba
The hypergeometric series should reduce to a polynomial of degree $N_r$.
Introducing the quantum number $n = 2 N_r + \beta$, the energy levels are
given by
$$
{\cal E} (\kappa) = ( n+1) \frac{1}{2} (( n+1) \kappa - \sqrt{\kappa^2 + 4}).
$$
The wavefunctions for the Schr\"odinger equation on a space with
constant curvature can be summarized as
\be
\Psi_{N_r , \beta} = C_{\kappa} r^{\beta} \, ( 1 + \kappa r^2 )^{\frac{1}{4}
- \frac{q (\kappa)}{4 \kappa}} \, _2F_1 (-N_r; b_{r} ; \beta +1; - \kappa r^2) 
\, e^{\pm i \beta \varphi}.
\label{eqV23}
\ba
In \eqref{eqV23}, $C_{\kappa}$ is a normalization constant. The energies $E$
are recovered by using \eqref{eqIV5} as
\be
E_n (\kappa) = \frac{\hbar \beta}{\sqrt{m}}
(n+1) \frac{1}{2} (( n+1) \, \kappa - \sqrt{ \kappa^2 + 4} ).
\label{eqV24}
\ba
The total energy \eqref{eqV24} is a linear function of the curvature
$\kappa$ and depends, as in the Euclidean case, on the combination
$n = 2 N_r + \beta$ of the quantum numbers $N_r$ and $\beta$.

\section{Summary}

The motion of the quantum free particle has been studied on
spherical and hyperbolic spaces using a curvature dependent approach.
The geometric approach was outlined at the start, and an important
step was the determination of three Killing vector fields by means
of Lie differentiation of the metric and then the associated Noether
symmetries. A Hilbert space was defined by calculating a measure
which is invariant under the same process of Lie differentiation
with respect to these Killing vector fields. It is worth noting again
that this measure is different from the measure that appeared in {\bf [10]}.
It is a bit reminiscent of the change in the path integral measure
in formulations of non-abelian gauge theories
which give rise to anomalies. Quantization of the three
Noether momenta as self-adjoint operators with respect to the
$\kappa$-dependent measure was carried out here, and the construction 
of the quantum Hamiltonian based on them in terms of the related
operators $\hat{P}_1$, $\hat{P}_2$ and $\hat{J}$ obtained from the 
Noether symmetries. Finally, it was found that the Schr\"odinger 
equation can be separated. This has led to a determination of
both the spectrum and the eigenfunctions of the quantum 
Hamiltonian for the choice of metric.

\section{References.}

\noindent
$[1]$ L. D. Landau and E. M. Lifshitz, Quantum Mechanics,
(Pergamon Press, Oxford, 1977).   \\
$[2]$ E. Prugovecki, Quantum Mechanics in Hilbert Space,
(Academic Press, New York, 1971).  \\
$[3]$ Z. F. Ezawa, Quantum Hall Effect, 3 rd. ed. 
(World Scientific, Singapore, 2013).  \\
$[4]$ M. Lakshmanan and S. Rajackar, Nonlinear Dynamics, Integrability,
Chaos and Patterns, (Springer-Verlag, Berlin, 2003).  \\
$[5]$ J. Sniatycki, Geometric Quantization and Quantum Mechanics,
(Springer-Verlag, New York, 1980).  \\
$[6]$ D. M. Xun, Q. H. Liu and X. M. Zhu, ``Quantum motion on a torus 
as a submanifold problem in a generalized Dirac's theory of second-class
constraints'', Ann. Phys., {\bf 338}, 123-133, (2013).  \\
$[7]$ J. F. Cari\~nena,  M. F. Re\~nada and M. Santander,
``The quantum free particle on spherical and hyperbolic spaces'',
J. Math. Phys., {\bf 52}, 072104, (2011).  \\
$[8]$ J. F. Cari\~nena, M. F. Re\~nada and M. Santander,
``The quantum harmonic oscillator on the sphere and the hyperbolic plane: 
$\kappa$-dependent formalism, polar coordinates and hypergeometric
functions'', J. Math. Phys., {\bf 48}, 102106, (2007).  \\
$[9]$ J. F. Cari\~nena, M. F. Ra\~nada and M. Santander, 
``A quantum exactly solvable non-linear oscillator with quasi-harmonic behavior'',
Ann. Phys., {\bf 322}, 434-459, (2007).   \\
$[10]$ P. Bracken, ``Motion on Constant Curvature Spaces and Quantization
Using Noether Symmetries'', arXiv:1406.2753v1, (2014).   \\
$[11]$ J. E. Marsden and T. S. Ratiu, Introduction to Mechanics and Symmetry,
(Springer-Verlag, New York, 1994).  \\
$[12]$ B. W. Char, K. O. Geddes, G. H. Gonnet, B. L. Leong,
M. B. Monagen and S. M. Watt, Maple V Library Manual (Springer, New York, 1994). \\
\end{document}